# Extreme nonlinear optical enhancement in chalcogenide glass fibers with deep-subwavelength metallic nanowires

Bora Ung[*] and Maksim Skorobogatiy

*Dept. of Engineering Physics, Ecole Polytechnique de Montreal, C.P. 6079, succ. Centre-ville, Montréal, Canada, H3C 3A7*
*Corresponding author: boraung@gmail.com*

**Abstract:** A nanostructured chalcogenide-metal optical fiber is proposed. This hybrid nanofiber is embedded with a periodic array of triangular-shaped deep-subwavelength metallic nanowires set up in a bowtie configuration. Our simulations show that the proposed nanostructured fiber supports a guided collective plasmonic mode enabling both subwavelength field confinement and extreme nonlinear light-matter interactions, much larger than a bare chalcogenide nanowire of comparable diameter. This is all achieved with less than 3% by volume of metal content.

In the last decade, different types of photonic crystal fibers and microstructured optical fibers have demonstrated exceptional control over the group velocity chromatic dispersion (GVD) [1]. More recently, advances in micro and nano-fabrication techniques have sparked vigorous research on microstructured optical fibers with subwavelength features, and using high refractive index compound glasses. These so-called *emerging waveguides* allow the exploration of new operation regimes where tight field confinement, enhanced light-matter nonlinear interactions and dispersion engineering can be combined to enable long interaction lengths inside nonlinear media [2]. In parallel, the merging of plasmonics with integrated optics has shown vast potential for sensing [3], and the transmission [4,5] and modulation of optical signals on the subwavelength scale [6-8]. Light guiding mediated by metallic nanowire arrays in optical fibers was also investigated [9-12], and their potential for nonlinear optical plasmonics was mentioned in [9] but not studied in detail.

In this paper, we present a new type of nonlinear metallo-dielectric nanostructured optical fiber (NOF): the chalcogenide fiber with deep-subwavelength metallic inclusions. We demonstrate that the extreme field intensities obtained at the sharp edges of the subwavelength metallic nanowires enable giant nonlinear optical enhancements to be achieved inside the chalcogenide glass. We show that modal propagation losses are comparable to that of classical surface plasmons polaritons (SPP) and thus only limited by the intrinsic absorption losses of the metal.

When fabricating a fiber comprising several dielectric rods (or capillaries) using the stack-and-draw procedure, the empty interstitial holes [Figs. 1(a) and 1(b)] that may occur between adjacent rods are usually treated as unwanted defects. Here we fill these nanovoids with metal such that we obtain a symmetrical periodic array of triangle-shaped metallic nanowires. Coincidentally, the triangular "bowtie" structure is known to be a very effective configuration for yielding strong field confinement in the nanogap between two apexes [13]. In the proposed fiber design, the intense local fields enabled by metallic bowtie nanowires enhance the nonlinear light-matter interaction within the chalcogenide glass matrix [Figs. 1(c) and 1(d)].

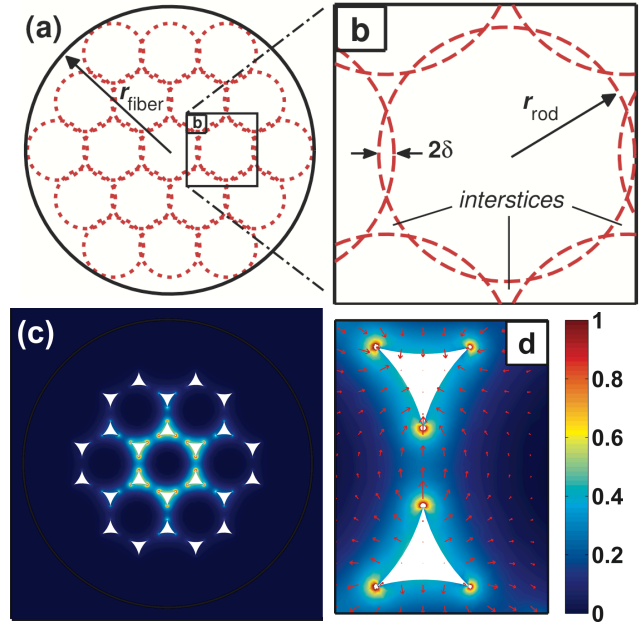

Fig. 1. (Color online) (a) Configuration of overlapping rods and interstices with (b) close-up view of the geometry. (c) $S_z$-flux distribution of the fundamental plasmonic supermode in the chalco-metallic nanostructured fiber ($r_{fiber}$ = 0.450 μm, $f_\delta$ = 0.03) at $\lambda$=3.0 μm, and (d) close-up view of the enhanced local fields (arrows denote vector $E$-fields) in a "bowtie" pair of nanowires.

Towards the fabrication of metal (or chalcogenide) nanowires embedded in a dielectric matrix, a practical method was recently demonstrated in [14] where the authors used a consecutive stack-and-draw technique to produce nanowires with extreme aspect ratios. A potential fabrication strategy of this NOF is to first stack identical circular glass rods in a triangular lattice configuration in the preform, and then allow the rods to overlap during the thermal softening and drawing process. The ensuing small air interstices are then filled by pumping molten metal at high pressure, as demonstrated in [9]. This interstice-filling approach allows the desirable edge-to-edge alignment of pairs of nanowires to be enforced right from the initial macroscopic preform. The wavelength of interest in the present study, $\lambda$=3.0 μm, is in the middle-infrared. The chosen metal is gold (Au) while the dielectric material is $As_2Se_3$ chalcogenide glass due to its wide bulk transmission window inside the mid-infrared (2 - 14 μm) and its large nonlinear index value: $n_2 = 1.1 \times 10^{-17}$ $m^2$/W.

To simplify analysis, the present investigation focuses on the case of $N$ = 2 layers of rods (of equal radiuses $r_{rod}$) with gold-filled interstices [Figs. 1(a)-(b)]. Nevertheless, the results and discussion presented thereafter provides the foundation for more complex chalco-metallic fibers ($N$ > 2). In our model the overlapping of rods is tuned by the "overlap half-distance" $\delta$ as defined in Fig. 1(b), and in practice this is accomplished by controlling the injected gas pressure when drawing the fiber. Here we use the convenient non-dimensional parameter $f_\delta = \delta/r_{rod}$ where $f_\delta = 0$ defines the case of tangent circles; while positive values $0 < f_\delta < f_{\delta,max}$ controls the overlapping of adjacent rods up to a maximal ratio $f_{\delta,max} = 0.1339$ (i.e. the limit where the area of the triangular-shaped interstice disappears). The total outer radius of the fiber is set as $r_{fiber} = 5r_{rod} - 3\delta$ such that the whole geometry can be specified using only $r_{fiber}$ and the "overlap factor" ($f_\delta$) which was kept fixed at 0.03 in this study, corresponding to 2.8% by volume of metal content.

Recently, there have been demonstrations of large nonlinear optical enhancement in chalcogenide nanowires [15]. Therefore as a benchmark, we compare below the optical properties of our hybrid NOF with the bare nanowire (i.e. rod-in-air). Using fully-vectorial finite-element simulations, we solved for the fundamental $HE_{11}$ mode guided in the bare $As_2Se_3$ nanowire and for the fundamental plasmonic supermode [Fig. 1(c)] guided in the metallo-dielectric NOF, at various values of fiber radius. By comparing the bare nanowire with a hybrid NOF of the same size ($r_{fiber}$ = 0.32 μm), one can appreciate the exceptionally strong transverse field confinement in chalco-metallic fibers [left inset of Fig. 2(a)]; whilst in the case of the nanowire there is very significant power leakage into the surroundings [right inset of Fig. 2(a)]. Figure 2(b) indicates that the chalco-metallic fiber consistently provides better field confinement over a chalcogenide nanowire of the same size. In particular, one achieves a nearly tenfold enhancement in field confinement for $r_{fiber} \geq 0.415$ μm. For $r_{fiber} < 0.415$ μm, while the effective mode area of a bare nanowire diverges to the completely unguided limit ($A_{eff} \rightarrow \infty$ for $r_{fiber} \rightarrow 0$), the mode area supported by a chalco-metallic fiber keeps shrinking with smaller radiuses of fiber. This nanoscale localization of light beyond the diffraction limit is made possible by the plasmonic guiding in the metallic nanowire array.

The full-vector calculations indicate that the hybrid $As_2Se_3$-gold NOF can theoretically yield nonlinearities over $1\times10^4$ $W^{-1}m^{-1}$ at $\lambda$=3.0 μm [Fig. 2(a)]. In particular, the nonlinearity reaches a peak of $\gamma$=286 $W^{-1}m^{-1}$ at $r_{fiber}$ = 0.415 μm for the bare nanowire; while at the same radius size the hybrid NOF provides $\gamma$=2.63×10$^4$ $W^{-1}m^{-1}$ corresponding to a nonlinear enhancement factor of 92, almost 2 orders of magnitude increase in nonlinearity at the given operation wavelength.

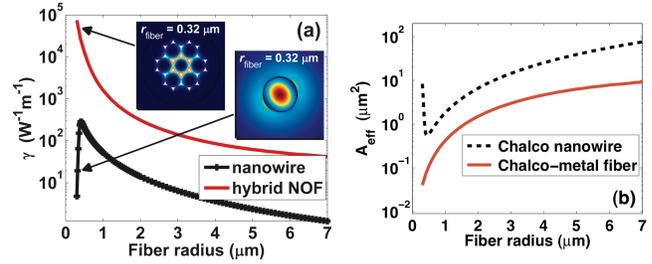

Fig. 2. (Color online) (a) Optical nonlinearity ($\gamma$), and (b) effective mode area ($A_{eff}$), of a bare chalcogenide nanowire and a chalco-metallic nanofiber at $\lambda$=3.0 μm as a function of the fiber radius.

The nonlinear enhancement factor in chalco-metallic fibers (compared to similar nanowires) is consistently greater than 20 across the investigated range of fiber radiuses. The effective area and nonlinear parameter were computed using their respective vectorial expressions [16] taking $S_z = (\vec{E} \times \vec{H}) \cdot \hat{z}$:

$$A_{eff} = \frac{\left|\int_{total} S_z \, dA\right|^2}{\int_{total} |S_z|^2 \, dA} \quad (1)$$

$$\gamma = k \frac{\varepsilon_0}{\mu_0} \cdot \frac{\int_{total} n^2(x,y) \cdot n_2(x,y) \cdot \left[2|\vec{E}|^4 + |\vec{E}^2|^2\right] \cdot dA}{3\left|\int_{total} S_z \cdot dA\right|^2} \quad (2)$$

The effective refractive index $Re(n_{eff})$ and effective propagation length ($L_{eff}$) at $\lambda$=3.0 μm of the fundamental guided mode in the chalcogenide nanowire and the chalco-metal NOF as a function of fiber radius are plotted in Fig. 3(a) and Fig. 3(b) respectively.

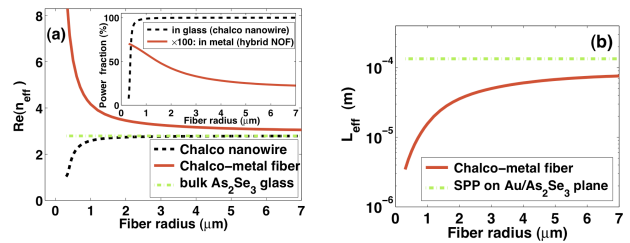

Fig. 3. (Color online) (a) Real effective refractive index ($n_{eff}$), and (b) effective propagation length ($L_{eff}$), of a bare chalcogenide nanowire and a chalco-metallic nanofiber at $\lambda$=3.0 μm as a function of the fiber radius. Inset of (a): fraction of power guided inside metallic regions for the chalco-metallic fiber (solid line); and in chalcogenide glass for the nanowire (dashed line). The $n_{eff}$ and $L_{eff}$ of a SPP on the gold - $As_2Se_3$ planar interface is shown in green dashed-dot line as reference, in Fig. (a) and (b) respectively.

As shown in the inset of Fig. 3(a), the fraction of guided power of the fundamental mode in the bare nanowire becomes more confined inside the chalcogenide glass matrix (and less in the air cladding) as the radius of the fiber increases. Consequently we observe in Fig. 3(a) for the nanowire: $n_{eff} \rightarrow n_{bulk}$ as $r_{fiber} \rightarrow \infty$, where $n_{bulk}$=2.789 at

$\lambda$=3.0 µm for As$_2$Se$_3$ chalcogenide glass. The reverse is also true in the unguided limit: the value of $n_{eff}$ reaches the refractive index of air cladding such that $n_{eff} \rightarrow n_{air} = 1$ as $r_{fiber} \rightarrow 0$. On the other hand for the hybrid NOF, the value of $n_{eff}$ increases with reduction of the fiber radius. The latter behavior can be explained by considering the fraction of power that propagates inside metallic regions [solid curve in inset of Fig. 3(a)]. One must first note that the geometrical proportions of the fiber's internal structure are preserved as the radius gets smaller. Hence for very small NOF radiuses ($r_{fiber}$ < 1 µm), the dimensions of the metallic nanowires become deeply subwavelength and comparable to the skin depth in metal such that a significant fraction of guided power overlaps with the strongly dispersive metallic regions.

The previous remark also explains the gradual lowering of the modal effective propagation length $L_{eff} = c/2\omega \, \text{Im}(n_{eff})$ in the hybrid NOF when $r_{fiber} \rightarrow 0$ [Fig. 3(b)]. Figure 3(b) indicates that the effective propagation length of the chalco-metallic NOF is comparable in magnitude to the propagation length of a planar SPP (dashed-dotted line) whose effective index is given by $n_{eff}^2 = [\varepsilon_m \varepsilon_d / (\varepsilon_m + \varepsilon_d)]$ where $\varepsilon_m$ and $\varepsilon_d$ denote the dielectric constants of the metal and adjacent dielectric. The latter remark highlights the fact that the guided mode of interest in the metallo-dielectric NOF is a plasmonic supermode created by the collective gap-plasmon modes supported by the array of bowtie nanowires. A notable feature here is that a reduction of $r_{fiber}$ from 5.0 µm to 0.5 µm, translates into lowering the $L_{eff}$ value of the hybrid NOF by a factor of 10 [Fig. 3(b)] but at the same time enhancing its nonlinearity by 200-fold [Fig. 2(a)]. To alleviate the large plasmonic losses, the incorporation of optical gain via dipolar dopants in the dielectric host (Ex: dye molecules, silicon nanocrystals and quantum dots) has demonstrated some success [5,17].

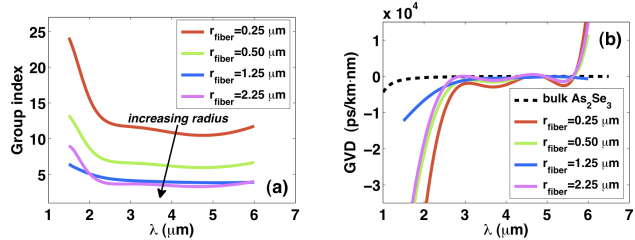

Fig. 4. (Color online) (a) Group index and (b) group velocity dispersion (ps/km/nm) for different values of radius of chalco-metallic NOFs as a function of wavelength. In (b) the GVD of bulk As$_2$Se$_3$ glass is shown (dashed curve) for reference.

The group index, defined as $n_g = n_{eff} + \omega(dn_{eff}/d\omega)$, for different values of fiber radius is plotted in Fig. 4(a) as a function of wavelength. We remark in Fig. 4(a) that the group index at $\lambda$=3.0 µm takes values from $n_g$ = 3.67 and $n_g$ = 11.68 with decreasing fiber size from $r_{fiber}$ = 2.25 µm to $r_{fiber}$ = 0.25 µm. Our simulations indicate that even larger slow-down factors (>20) could be attained at shorter near-infrared wavelengths where tighter field confinement is achieved. The group velocity chromatic dispersion (GVD) was computed using the definition $D = -\lambda/c(d^2 n_{eff}/d\lambda^2)$. Figure 4(b) shows that via appropriate choice of the geometrical parameters, the chalco-metallic NOF enables engineering of the GVD to create a zero-dispersion point (ZDP) at the $\lambda$=3.0 µm operation wavelength (magenta curve), or even a ZDP at $\lambda$=5.0 µm along with a flattened dispersion profile (blue curve) thus enabling larger effective bandwidths to be achieved. We also note some regions where the GVD is engineered to be anomalous, thus allowing nonlinear solitonic effects. Moreover, the ultrafast response (<50 fs) of the third-order nonlinearity in chalcogenide-based waveguides [18] indicates that the proposed NOF can potentially achieve broader bandwidths than all-optical processing devices operating on a resonant-based nonlinearity; and without free-carrier effects present in other materials such as silicon. These examples clearly suggest that the proposed hybrid NOF design is not only useful for extreme nonlinear optical interactions but also for slow light applications with an engineered chromatic dispersion profile.

In conclusion, a novel type of nanostructured optical fiber has been proposed. Our theoretical calculations demonstrate that the hybrid chalcogenide-metal nanofiber provides a platform for achieving nanoscale mode area nonlinear light-matter interactions, and also constitutes an alternative path for investigating slow-light applications. Therefore the proposed chalco-metallic fibers are relevant to the study of extreme nonlinear light-matter interactions and slow-light guiding systems for all-optical signal processing and highly-integrated nanophotonic devices in general. It also broadens the scope of both conventional and exotic physical phenomenae which can be conveniently studied through the use of micro(-nano)structured optical fibers.